\documentclass{amsart}
\usepackage{amsthm}                 
\usepackage{amsmath}                
\usepackage{amsfonts,amssymb}       
\usepackage{graphicx}               
\usepackage{float}                  
\usepackage{rotating}               
\restylefloat{figure}               
\usepackage{mathrsfs}               
\usepackage{fancyhdr}               
\usepackage{hyperref}               
\usepackage[numbers]{natbib}
\theoremstyle{plain}
\newtheorem{lem}{Lemma}

\newtheorem{thm}[lem]{Theorem}
\newtheorem{cor}[lem]{Corollary}

\theoremstyle{definition}
\newtheorem{defn}[lem]{Definition}
\newtheorem{exmp}[lem]{Example}

\newcommand{\R}{\mathbb R}

\newcommand{\Z}{\mathbb Z}

\newcommand{\Diff}{\mbox{\rm Diff}}
\newcommand{\Vect}{\mbox{\rm Vect}}

\newcommand{\dx}{\,\text{\rm d}x}
\renewcommand{\d}{\,\text{\rm d}}

\newcommand{\g}{\mathfrak{g}}

\newcommand{\ad}{\mbox{\rm ad}}

\renewcommand{\S}{\mathbb S}
\renewcommand{\phi}{\varphi}

\newcommand{\ska}[2]{\left\langle #1,#2\right\rangle}

\newcommand{\bea}{\begin{eqnarray}}
\newcommand{\eea}{\end{eqnarray}}
\renewcommand{\Phi}{\phi}

\newcommand{\eins}{\mathbf{1}}
\begin{document}
\title{The periodic $\mu$-$b$-equation and Euler equations on the circle}
\author{Martin Kohlmann}
\address{Institute for Applied Mathematics, University of Hannover, D-30167 Hannover, Germany}
\email{kohlmann@ifam.uni-hannover.de}
\begin{abstract} In this paper, we study the $\mu$-variant of the periodic
$b$-equation and show that this equation can be realized
as a metric Euler equation on the Lie group $\Diff^{\infty}(\S)$
if and only if $b=2$ (for which it becomes the $\mu$-Camassa-Holm equation). In
this case, the inertia operator generating the metric on
$\Diff^{\infty}(\S)$ is given by $L=\mu-\partial_x^2$. In
contrast, the $\mu$-Degasperis-Procesi equation (obtained for $b=3$) is not a metric
Euler equation on $\Diff^{\infty}(\S)$ for any regular inertia
operator $A\in\mathcal
L_{\text{is}}^{\text{sym}}(C^{\infty}(\S))$. The paper generalizes
some recent results of \cite{EK,ES10,Ko09}.
\end{abstract}
\maketitle
For the mathematical modelling of fluids, the so-called family of
$b$-equations
\bea\label{b}m_t=-(m_xu+bmu_x),\quad m=u-u_{xx},\eea
attracted a considerable amount of attention in recent years.
Here, $b$ stands for a real parameter, \cite{EY}. Each of these
equations models the unidirectional irrotational free surface flow
of a shallow layer of an inviscid fluid moving under the influence
of gravity over a flat bed. In this model $u(t,x)$ represents the
wave's height at time $t\geq 0$ and position $x$ above the flat
bottom. If the wave profile is assumed to be periodic,
$x\in\S\simeq\R/\Z$; otherwise $x\in\R$. For further details
concerning the hydrodynamical relevance we refer to
\cite{CL,Iv,Jo}. As shown in \cite{DP,HW,Iva,MN}, the $b$-equation
is asymptotically integrable which is a necessary condition for
complete integrability, but only for $b=2$ and $b=3$ for which it
becomes the Camassa-Holm (CH) equation
$$u_t-u_{txx}+3uu_x-2u_xu_{xx}-uu_{xxx}=0$$
and the Degasperis-Procesi (DP) equation
$$u_t-u_{txx}+4uu_x-3u_xu_{xx}-uu_{xxx}=0$$
respectively. The Cauchy problems for CH and DP have been studied
in detail: For the CH, there are global strong as well as global
weak solutions. In addition, CH allows for finite time blow-up
solutions which can be interpreted as breaking waves and there are
no shock waves (see, e.g., \cite{CE98,CE98'',CE2}). Some recent global
well-posedness results for strong and weak solutions, precise
blow-up scenarios and wave breaking for the DP are discussed in
\cite{ELY06,ELY07,Yi03,Yi,Z}.
\\
\indent Besides the various common properties of the CH and the DP
there are also significant differences to report on, e.g., when
studying geometric aspects of the family (\ref{b}). The periodic
equation (\ref{b}) reexpresses a geodesic flow on the group
$\Diff^{\infty}(\S)$ of smooth and orientation preserving
diffeomorphisms of the circle, cf.\ \cite{EK}.\ If $b=2$, the
geodesic flow corresponds to the right-invariant metric induced by
the inertia operator $1-\partial_x^2$ whereas for $b\neq 2$,
equation (\ref{b}) can only be realized as a non-metric Euler
equation, i.e., as geodesic flow with respect to a linear
connection which is not Riemannian in the sense that it is
compatible with a right-invariant metric, cf.\
\cite{CoKo,CoKo2,ES10,Ko09}.

The idea of studying Euler's equations of motion for perfect
(i.e., incompressible, homogeneous and inviscid) fluids
as a geodesic flow on a certain diffeomorphism
group goes back to \cite{A,EM} and in a recent work \cite{EK},
Escher and Kolev show that the theory is also valid for the
general $b$-equation.

In this paper, we are interested in the following variant
of the periodic family (\ref{b}). Let $\mu(u)=\int_\S u(t,x)\dx$
and $m=\mu(u)-u_{xx}$ in (\ref{b}) to obtain the family of
$\mu$-$b$-equations, cf.~\cite{LMT}. The study of the
$\mu$-variant of (\ref{b}) is motivated by the following key
observation: Letting $m=-\partial_x^2u$, equation (\ref{b}) for
$b=2$ becomes the Hunter-Saxton (HS) equation, cf.~\cite{HS91}, which possesses
various interesting geometric properties, cf.~\cite{Len1,Len2},
whereas the choice $m=(1-\partial_x^2)u$ leads to the CH as
explained above. In the search for integrable equations that are
obtained by a perturbation of $-\partial_x^2$, the
$\mu$-$b$-equation has been introduced and it could be shown that
it behaves quite similarly to the $b$-equation; cf.~\cite{LMT}
where the authors discuss local and global
well-posedness as well as finite
time blow-up and peakons. Peakons are peculiar wave forms:
they are travelling wave solutions which are smooth except at their
crests; the lateral tangents exist, are symmetric but different.
Such wave forms are known to characterize the steady water waves of
greatest height, \cite{C06,CE07,T96}, and were first shown
to arise for the CH in \cite{CH93}.\\
\indent The goal of this paper is to extend the work done in
\cite{ES10} to the family of $\mu$-$b$-equations. Our main result
is that the periodic $\mu$-$b$-equation can be realized as a
metric Euler equation on $\Diff^{\infty}(\S)$ if and only if
$b=2$, for which it becomes the $\mu$CH equation. The corresponding regular
inertia operator is $\mu-\partial_x^2$. Before we give a proof, we
begin with some introductory remarks about Euler equations on
$\Diff^{\infty}(\S)$. In a first step, we comment on the operator
$\mu-\partial_x^2$.
\lem The bilinear map
$$\ska{\cdot}{\cdot}_\mu\colon C^{\infty}(\S)\times C^{\infty}(\S)\to\R,\quad \ska{u}{v}_\mu=\mu(u)\mu(v)+\int_\S u_x(x)v_x(x)\dx$$
defines an inner product on $C^{\infty}(\S)$.
\proof Clearly, $\ska{\cdot}{\cdot}_\mu$ is a symmetric bilinear form and $\ska{u}{u}_\mu\geq 0$. If $u\in C^{\infty}(\S)$ satisfies $\ska{u}{u}_\mu=0$,
then $u_x=0$ on $\S$ and hence $u$ is constant. The fact that $\mu(u)=0$ implies $u=0$.\endproof
We obtain a right-invariant metric on the Lie group
$G=\Diff^{\infty}(\S)$ by defining the inner product
$\ska{\cdot}{\cdot}_\mu$ on the Lie algebra $\g\simeq\Vect^{\infty}(\S)\simeq
C^{\infty}(\S)$ of smooth vector fields on $\S$ and transporting $\ska{\cdot}{\cdot}_\mu$ to any
tangent space of $G$ by using right translations, i.e., if
$R_\phi:G\to G$ denotes the map sending $\psi$ to $\psi\circ\phi$,
then
$$\ska{u}{v}_{\mu;\phi}=\ska{D_\phi R_{\phi^{-1}}u}{D_\phi R_{\phi^{-1}}v}_\mu,$$
for all $u,v\in T_\phi G$.
Observe that $\ska{\cdot}{\cdot}_\mu$ can be expressed in terms of
the symmetric linear operator $L\colon\g\to\g'$ defined by
$L=\mu-\partial_x^2$, i.e.,
$$\ska{u}{v}_\mu=\ska{Lu}{v}=\ska{Lv}{u},\quad u,v\in C^{\infty}(\S),$$
where $\ska{\cdot}{\cdot}$ denotes the duality pairing on
$\g'\times\g$.
\defn Each symmetric isomorphism $A\colon\g\to\g'$ is called an \emph{inertia operator} on $G$.
The corresponding right-invariant metric on $G$ induced by $A$ is denoted by $\rho_A$.\enddefn
Let $A$ be an inertia operator on $G$. We denote the Lie bracket
on $\g$ by $[\cdot,\cdot]$ and write $(\ad_u)^*$ for the adjoint
with respect to $\rho_A$ of the natural action of $\g$ on itself
given by $\ad_u:\g\to\g$, $v\mapsto[u,v]$. Let
$$B(u,v)=\frac{1}{2}\left[(\ad_u)^*v+(\ad_v)^*u\right].$$
We define a right-invariant linear connection on $G$ via
\bea\label{connection}\nabla_uv=\frac{1}{2}[u,v]+B(u,v),\quad u,v\in C^{\infty}(\S).\eea
As explained in \cite{EK,ES10}, we have the
following theorem.
\thm A smooth curve $g(t)$ on the Lie group $G=\Diff^{\infty}(\S)$ is a geodesic for
the right-invariant linear connection defined by
\text{\rm(\ref{connection})} if and only if its Eulerian velocity
$u(t)=D_{g(t)}R_{g^{-1}(t)}g'(t)$ satisfies the Euler equation
\bea\label{Euler}u_t=-B(u,u).\eea
\endthm\rm
Observe that the topological dual space of
$\Vect^{\infty}(\S)\simeq C^{\infty}(\S)$ is given by the
distributions $\Vect'(\S)$ on $\S$. In order to get a convenient
representation of the \emph{Christoffel operator} $B$ we restrict
ourselves to $\Vect^*(\S)$, the set of all regular distributions
which can be represented by smooth densities, i.e.,
$T\in\Vect^*(\S)$ if and only if there is a $\sigma\in
C^{\infty}(\S)$ such that
$$T(\phi)=\int_\S\sigma(x)\phi(x)\dx,\quad\forall\phi\in C^{\infty}(\S).$$
By means of the Riesz representation theorem we may identify
$\Vect^*(\S)\simeq C^{\infty}(\S)$. This motivates the following
definition.
\defn Let $\mathcal L_{\text{is}}^{\text{sym}}(C^{\infty}(\S))$ denote the set of all continuous isomorphisms on $C^{\infty}(\S)$
which are symmetric with respect to the $L_2$ inner product. Each
$A\in\mathcal L_{\text{is}}^{\text{sym}}(C^{\infty}(\S))$ is
called a \emph{regular inertia operator} on
$\Diff^{\infty}(\S)$.\enddefn
The following lemma establishes that the operator $L$ belongs to
the above defined class of regular inertia operators.
\lem The operator $L$ is a regular inertia operator on $\Diff^{\infty}(\S)$.
\proof One checks that applying $L$ to
\bea&&\left(\frac{1}{2}x^2-\frac{1}{2}x+
\frac{13}{12}\right)\int_0^1u(a)\d a+\left(x-\frac{1}{2}\right)
\int_0^1\int_0^au(b)\d b\d a\nonumber\\
&&\qquad-\int_0^x\int_0^au(b)\d b\d
a+\int_0^1\int_0^a\int_0^bu(c)\d c\d b\d a\nonumber\eea
gives back the function $u$. It is easy to see that if $u\in C^{\infty}(\S)$, then its
pre-image also belongs to $C^{\infty}(\S)$. Assume that $Lu=0$ for
$u\in C^{\infty}(\S)$. We thus can find constants $c,d\in\R$ such
that $u=\frac{1}{2}\mu(u)x^2+cx+d$. Since $u$ is periodic, $c=0$
and $\mu(u)=0$ and thus also $d=0$. Clearly,
$L:C^{\infty}(\S)\to C^{\infty}(\S)$ is bicontinuous.
\endproof
A proof of the following theorem can be found in \cite{ES10}.
\thm Given $A\in\mathcal L_{\text{\rm is}}^{\text{\rm
sym}}(C^{\infty}(\S))$, the Christoffel operator
$B=\frac{1}{2}[(\ad_u^*)v+(\ad_v^*)u]$ has the form
$$B(u,v)=\frac{1}{2}A^{-1}\left[2(Au)v_x+2(Av)u_x+u(Av)_x+v(Au)_x\right],$$
for all $u,v\in C^{\infty}(\S)$.
\endthm\rm
It may be instructive to discuss the following paradigmatic
examples.
\exmp\label{exmp}
Let $\lambda\in[0,1]$ and let $A$ be the inertia operator for the
equation $m_t=-(m_xu+2u_xm)$.
\begin{enumerate}
\item The choice $A=-\partial_x^2$ yields
$B(u,u)=-A^{-1}(2u_xu_{xx}+uu_{xxx})$ and $u_t=-B(u,u)$ is the
Hunter-Saxton equation
$$u_{txx}+2u_xu_{xx}+uu_{xxx}=0.$$
\item We choose $A=1-\lambda\partial_x^2$. If $\lambda=0$, the
equation $m_t=-(m_xu+2u_xm)$ becomes the periodic inviscid Burgers
equation $u_t+B(u,u)=u_t+3uu_x=0$. For $\lambda\neq 0$, we obtain
$$u_t+B(u,u)=u_t+3uu_x-\lambda(2u_xu_{xx}+uu_{xxx}+u_{txx})=0,$$
a $1$-parameter family of Camassa-Holm equations.
\item Choosing $A=\mu-\partial_x^2$, we arrive at the $\mu$CH
equation
$$\mu(u_t)-u_{txx}+2\mu(u)u_x=2u_xu_{xx}+uu_{xxx},$$
which is also called $\mu$HS in the literature, cf. \cite{KLM08}.
\end{enumerate}\endexmp\rm
Each regular inertia operator induces a metric Euler equation on
$\Diff^{\infty}(\S)$. We now consider the question for which
$b\in\R$ there is a regular inertia operator such that the
$\mu$-$b$-equation is the corresponding Euler equation on
$\Diff^{\infty}(\S)$. Example \ref{exmp} shows that, for $b=2$,
the operator $L\in\mathcal
L_{\text{is}}^{\text{sym}}(C^{\infty}(\S))$ induces the $\mu$CH.
Our goal is to show that this works only for $b=2$, and our main theorem
reads as follows.
\thm Let $b\in\R$ be given and suppose that there is a regular
inertia operator $A\in\mathcal L_{\text{\rm is}}^{\text{\rm
sym}}(C^{\infty}(\S))$ such that the $\mu$-b-equation
$$m_t=-(m_xu+bmu_x),\quad m=\mu(u)-u_{xx},$$
is the Euler equation on $\Diff^{\infty}(\S)$ with respect to $\rho_A$. Then $b=2$ and $A=L$.\endthm
\proof We assume that, for given $b\in\R$ and $A\in\mathcal
L_{\text{is}}^{\text{sym}}(C^{\infty}(\S))$, the
$\mu$-$b$-equation is the Euler equation on the circle
diffeomorphisms with respect to $\rho_A$. Then
$$u_t=-A^{-1}((Au)_xu+2(Au)u_x)$$
and the $\mu$-$b$-equation can be written as
$$(Lu)_t=-((Lu)_xu+b(Lu)u_x).$$
Using that $(Lu)_t=Lu_t$ and resolving both equations with respect
to $u_t$ we get that
\bea\label{7}A^{-1}\left(2(Au)u_x+u(Au)_x\right)=L^{-1}\left(b(Lu)u_x+u(Lu)_x\right),\eea
for $u\in C^{\infty}(\S)$. Denote by $\eins$ the constant function
with value 1. If we set $u=\eins$ in (\ref{7}), then
$A^{-1}(\eins(A\eins)_x)=0$ and hence $(A\eins)_x=0$, i.e.,
$A\eins=c\eins$. Scaling (\ref{7}) shows that we may assume $c=1$.
Replacing $u$ by $u+\lambda$ in (\ref{7}) and scaling with
$\lambda^{-1}$, we get on the left-hand side
\bea&&\frac{1}{\lambda}A^{-1}\big(2(A(u+\lambda))(u+\lambda)_x+(u+\lambda)(A(u+\lambda))_x\big)\nonumber\\
&&\quad=\frac{1}{\lambda}A^{-1}\big(2((Au)+\lambda)u_x+(u+\lambda)(Au)_x\big)\nonumber\\
&&\quad=A^{-1}\left(\frac{2(Au)u_x+u(Au)_x}{\lambda}+2u_x+(Au)_x\right)\nonumber\\
&&\quad\to A^{-1}(2u_x+(Au)_x),\,\lambda\to\infty,\nonumber\eea
and a similar computation for the right-hand side gives
\bea&&\frac{1}{\lambda}L^{-1}\big(b(L(u+\lambda))(u+\lambda)_x+(u+\lambda)(L(u+\lambda))_x\big)\nonumber\\
&&\quad\to L^{-1}(bu_x+(Lu)_x),\,\lambda\to\infty.\nonumber\eea
We obtain
\bea\label{8}A^{-1}\left(2u_x+(Au)_x\right)=L^{-1}(bu_x+(Lu)_x).\eea
We now consider the Fourier basis functions $u_n=e^{\text{i}nx}$
for $n\in 2\pi\Z\backslash\{0\}$ and have $Lu_n=n^2u_n$ and
$$L^{-1}(b(u_n)_x+(Lu_n)_x)=\text{i}\alpha_nu_n,\quad\alpha_n=\frac{b}{n}+n.$$
Next, we apply $A$ to (\ref{8}) with $u=u_n$ and see that
$$2\text{i}nu_n+(Au_n)_x=\text{i}\alpha_n(Au_n).$$
Therefore $v_n:=Au_n$ solves the ordinary differential equation
\bea\label{9}v'-\text{i}\alpha_nv=-2\text{i}nu_n.\eea
If $b=0$, then $\alpha_n=n$ and hence the general solution of (\ref{9}) is
$$v(x)=(c-2\text{i}nx)u_n,\quad c\in\R,$$
which is not periodic for any $c\in\R$. Hence $b\neq 0$ and there are numbers $\gamma_n$ so that
$$v_n=Au_n=\gamma_n e^{\text{i}\alpha_nx}+\beta_nu_n,\quad\beta_n=\frac{2}{b}n^2.$$
We first discuss the case $\gamma_n=0$ for all $n$ and show that
$\gamma_p\neq 0$ for some $p\in 2\pi\Z\backslash\{0\}$ is not
possible. If all $\gamma_n$ vanish, then $Au_n=\beta_nu_n$ and $A$
is a Fourier multiplication operator; in particular $A$ commutes
with $L$. Therefore (\ref{7}) with $u=u_n$ is equivalent to
$$L(2(Au_n)(u_n)_x+u_n(Au_n)_x)=A(b(Lu_n)(u_n)_x+u_n(Lu_n)_x)$$
and by direct computation
$$12\text{i}n^3\beta_nu_{2n}=\text{i}(b+1)n^3\beta_{2n}u_{2n}.$$
Inserting $\beta_{n}=2n^2/b$ we see that $b=2$ and $\beta_n=n^2$.
Therefore $A=L$. Assume that there is $p\in 2\pi\Z\backslash\{0\}$
with $\gamma_p\neq 0$. Since $v_p=Au_p$ is periodic, $\alpha_p\in
2\pi\Z$ and hence $b=kp$ for some $k\in 2\pi\Z\backslash\{0\}$. Let
$\alpha_p=m$. If $m=p$, then $b=0$ which is impossible. We thus
have $\ska{u_m}{u_p}=0$ and
$$\ska{Au_p}{u_m}=\ska{\gamma_pe^{\text{i}mx}}{u_m}=\gamma_p.$$
The symmetry of $A$ yields
$$\gamma_p=\ska{Au_p}{u_m}=\ska{u_p}{Au_m}=\overline{\gamma_m}\ska{u_p}{e^{\text{i}\alpha_mx}}.$$
Since $\gamma_p\neq 0$, $\gamma_m$ is non-zero and periodicity
implies $\alpha_m\in 2\pi\Z$. More precisely, $\alpha_m=p$ since
otherwise $\ska{u_p}{e^{\text{i}\alpha_mx}}=0=\gamma_p$. Using $b=kp$ and
the definition of $\alpha_p$, we see that $m=\alpha_p=k+p$.
Furthermore,
$$p(k+p)=\alpha_m(k+p)=\alpha_{k+p}(k+p)=kp+(k+p)^2$$
and hence $0=k^2+2pk$. Since $k\neq 0$, it follows that $k=-2p$
and hence $b=-2p^2$. We get $\alpha_p=-p$ and observe that
$\gamma_n=0$ for all $n\notin\{p,-p\}$, since otherwise repeating
the above calculations would yield $b=-2n^2$ contradicting
$b=-2p^2$. Inserting $u=u_p$ in (\ref{7}) shows that
$$\text{i}p\gamma_p\eins-\frac{3\text{i}p}{\beta_{2p}}u_{2p}=\text{i}p^3(b+1)\frac{u_{2p}}{4p^2};$$
here we have used that $Au_p=\gamma_p/u_p+\beta_pu_p$,
$\beta_p=-1$ and $A^{-1}u_{2p}=u_{2p}/\beta_{2p}$, since $2p$ does
not coincide with $\pm p$ and hence $\gamma_{2p}=0$. It follows
that $p\gamma_p=0$ in contradiction to $p,\gamma_p\neq 0$.
\endproof
\cor The $\mu$\text{DP} equation on the circle
$$m_t=-(m_xu+3mu_x),\quad m=\mu(u)-u_{xx},$$
cannot be realized as a metric Euler equation for any
$A\in\mathcal L_{\text{\rm is}}^{\text{\rm sym}}(C^{\infty}(\S))$.
\endcor\rm
\end{document}